\documentclass[aps,pra,twocolumn]{revtex4}
\usepackage{graphicx}
\usepackage{ulem}

\begin{document}
\title{Optical Synthesis of Terahertz and Millimeter-Wave Frequencies with Discrete 
Mode Diode Lasers}

\author{Stephen O'Brien}
\email{stephen.obrien@tyndall.ie}
\author{Simon Osborne}
\author{David Bitauld}
\author{Nicola Brandonisio}
\author{Andreas Amann}
\affiliation{Tyndall National Institute, University College Cork, Lee Maltings, Cork, Ireland}
\author{Richard Phelan}
\author{Brian Kelly}
\author{James O'Gorman}
\affiliation{Eblana Photonics, Trinity College Enterprise Centre, Pearse Street, Dublin 2, Ireland}

\begin{abstract}

It is shown that optical synthesis of terahertz and millimeter-wave frequencies can be 
achieved using two-mode and mode-locked discrete mode diode lasers. These edge-emitting 
devices incorporate a spatially varying refractive index profile which is designed
according to the spectral output desired of the laser. We first demonstrate a 
device which supports two primary modes simultaneously with high spectral purity. 
In this case sinusoidal modulation of the optical intensity at terahertz frequencies 
can be obtained. Cross saturation of the material gain in quantum well lasers prevents
simultaneous lasing of two modes with spacings in the millimeter-wave region. We show 
finally that by mode-locking of devices that are designed to support a minimal set of four 
primary modes, we obtain a sinusoidal modulation of the optical intensity 
in this frequency region. 

\end{abstract}


\maketitle

\section{Introduction}

Synthesis of terahertz (THz) and millimeter-wave (mmWave) 
frequencies using optoelectronic means is currently the subject of significant
interest. Generation of these frequencies in the optical domain can be 
much less complex than with electrical schemes and also enables signal 
distribution with the large bandwidth and low losses of optical fibers. 
Interesting applications are expected in diverse areas including 
high-bandwidth wireless communication systems, THz imaging and spectroscopy, 
and radio-astronomy \cite{siegel_02, yao_09}. 

The most direct route to optical synthesis of these frequencies is simply to
mix the outputs of two detuned single mode lasers \cite{mcintosh_95}. This 
technique is simple to implement and provides very wide tunability. 
However, because the outputs of the two sources are not correlated, broadening 
of the linewidth of the generated difference frequency is observed. 
For many applications, an active feedback mechanism such an optical phase 
locked loop is necessary to improve the linewidth \cite{steele_83, friederich_10}. 

An alternative for many applications is to electronically modulate the output of
a continuous wave single mode laser \cite{hirata_03, fukushima_03}. Electronics 
based techniques can combine frequency tunability and narrow linewidth but it 
is necessary to filter the modes of interest from the intensity modulated 
spectrum. In addition, higher frequency electronic modulation schemes are 
correspondingly more complex and expensive to implement. 

Two-mode diode lasers represent an interesting alternative approach to the 
problem on account of their compactness and simplicity. While monolithic 
two-mode devices do not generally lead to continuous and wide tunability of 
the difference frequency, there may be significant advantages where the 
frequency of operation is fixed for a particular application. 
Because the modes share a common cavity, the degradation of the 
linewidth of the beatnote due to environmental and other 
factors is expected to be less than in the case of two independent 
single mode lasers \cite{tani_05}. Many examples of semiconductor lasers 
that support two modes with a large frequency spacing have been described 
in the literature. These include external cavity lasers \cite{park_04}, 
and devices based on distributed Bragg reflections \cite{iio_95, matsui_99, 
roh_00} and compound cavity designs \cite{avrutin_02, kim_09}.

Here we demonstrate a powerful approach to mode selection in a Fabry-Perot 
semiconductor laser that enables optical synthesis of frequencies 
ranging from THz to the mmWave region. Depending on the frequency required, 
two-mode or mode-locked diode lasers that support a minimal set of four 
pre-selected modes can be used. Our approach relies on a distributed 
reflection mechanism that is designed as a perturbation of the Fabry-Perot 
mode spectrum. The cavity geometry is derived directly from an inverse 
problem solution that has enabled us to design single-mode \cite{obrien_05} 
and two-mode lasers \cite{obrien_06a} with excellent spectral purity. 
Single mode devices of this kind were also shown to have very narrow 
linewidth emission under normal drive conditions \cite{kelly_07}. 

This paper is organised as follows: In section II, we demonstrate that two-mode 
quantum-well lasers provide a direct route to optical synthesis of THz frequencies. 
The stability properties of two-mode devices as a function of the primary mode 
spacing are also described. It is shown that for primary mode spacings in the 
mmWave region, cross saturation of the material gain leads to wavelength bistability 
of the primary modes. In section III, we describe how mode-locking of a device 
designed to support a minimal set of four primary modes can be used to overcome the problem of 
wavelength bistability. Finally, prospects for obtaining narrow linewidth THz 
and mmWave signals using discrete mode devices are discussed.  

\section{Stability properties of two-mode devices as a function of primary mode spacing} 
 
We have shown that a set of self-consistent equations for the lasing modes can 
be found by making an expansion about the cavity resonance condition in a 
Fabry-Perot laser \cite{obrien_06a}. The effective index along the Fabry-Perot 
cavity is perturbed by $N$ additional features that are described by an index 
step $\Delta n$. A first order scattering approximation allows Fourier analysis 
to be used to make a direct connection between the index profile in real space 
and the threshold gain modulation in wavenumber space. Here we will present 
measurements of fabricated devices where the index step features are slotted regions 
etched into the ridge waveguide of the laser. All of the devices considered 
are multi-quantum well InP/InGaAlAs devices with peak emission ranging from 
1.3 $\mu$m to 1.5 $\mu$m. 

\begin{figure}
\centering
\includegraphics[width=8.0cm]{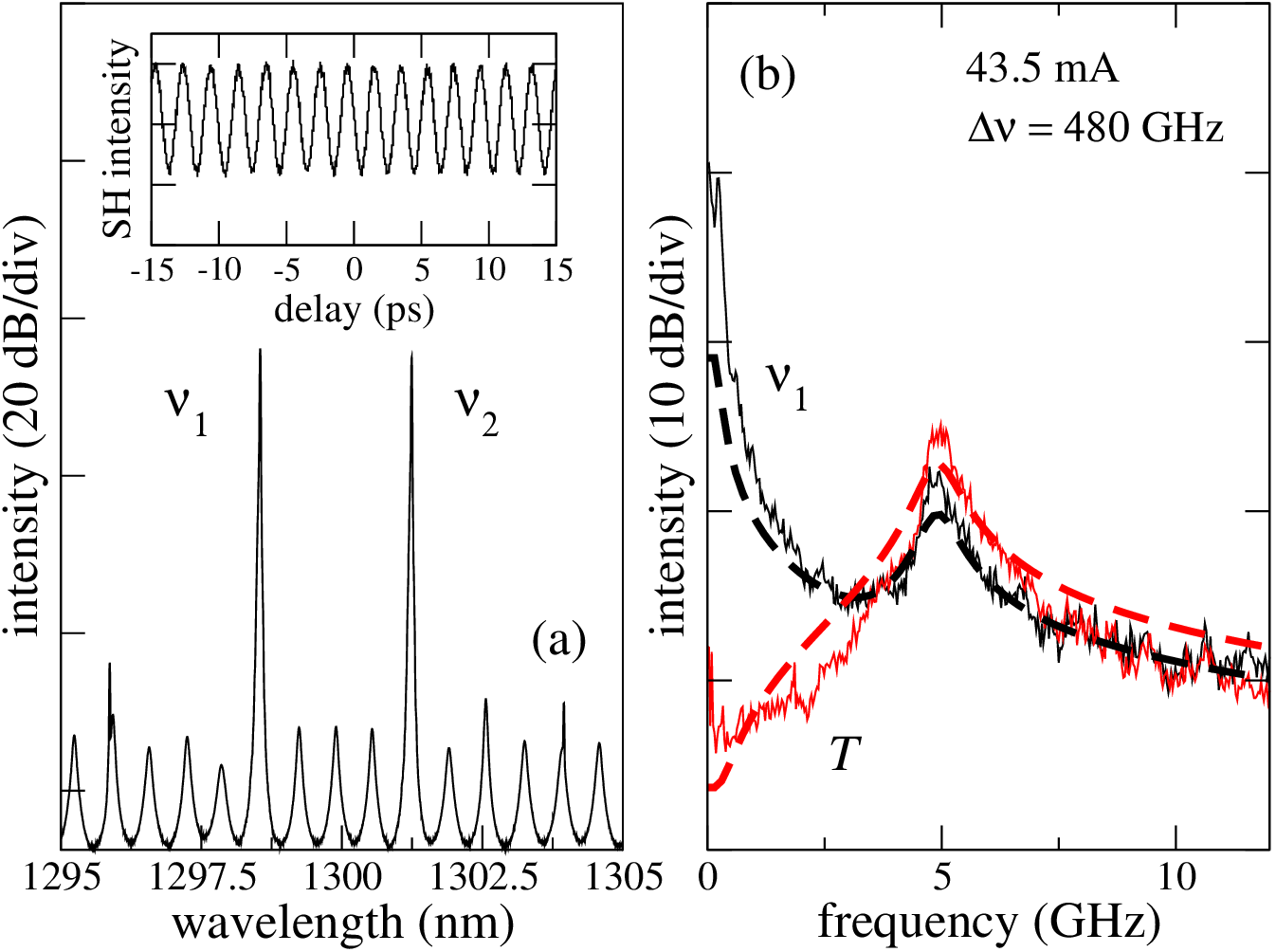}
\caption{\label{2mspc4A} (a) Optical spectrum of the device at a device current of 43.5 mA
(two-color point). Inset: Background free intensity autocorrelation measurement showing 
mode beating at 480 GHz. (b) Power spectra of one of the primary modes, $\nu_1$, 
and of the total intensity, $T$, measured at the two-color point. The dashed lines are  
theoretical calculations obtained using equation \ref{eq:6}.}
\end{figure}

Measurements of a two-mode ridge waveguide Fabry-Perot laser with a peak emission 
near 1.3 $\mu$m are shown in Fig. \ref{2mspc4A} (a). Details of the inverse problem
solution and of the design of two-mode diode lasers can be found in \cite{obrien_06a}. 
In this case the device length is 350 $\mu$m, and we have chosen a primary mode spacing of 
four modes which determines a frequency separation of the primary modes of 480 GHz. 
Generally, one finds that as the current is increased from below threshold the primary 
mode on the short wavelength side reaches threshold first and as the current is increased 
further the peak power shifts across the primary mode spacing to the long wavelength 
side. The spectrum shown in Fig. \ref{2mspc4A} (a) was obtained at 
a device current of 43.5 mA. At the value of the device current shown the time averaged 
optical power in each of the primary modes is approximately equal. We refer to this point
in the light-current relationship as the two-color point. One can see that excellent 
mode selectivity is achieved in this case with the primary mode intensity exceeding 
the background modes by a factor of 40 dB. 

The spectral purity observed in Fig. \ref{2mspc4A} (a) is limited in this case by power 
transfer to four-wave mixing sidebands. The presence of these sidebands is significant as 
they indicate simultaneous lasing of the primary modes. This is confirmed by the intensity 
autocorrelation measurement shown in the inset of Fig. \ref{2mspc4A} (a), which confirms a 
sinusoidal modulation of the intensity output at the primary mode frequency separation of 
480 GHz. Power spectra of one of the primary modes and of the total 
intensity are shown in Fig. \ref{2mspc4A} (b). One can see that there is significant 
low-frequency antiphase noise in the spectrum of the primary mode, which can be attributed 
to the global coupling of modes by the carrier density in the device. 

These characteristic features of the observed power spectra can be reproduced on the basis 
of a rate-equation model for the dynamics of the electric fields of each mode $E_1$ and
$E_2$ and for the carrier density $N$. To account for the spontaneous emission noise we add 
Gaussian white noise terms $\eta_1(t)$ and $\eta_2(t)$ to the dynamical equations for the fields 
and arrive at a system of stochastic differential equations of the form
\begin{eqnarray}
&&\dot{E}_{1} = \textstyle{\frac{1}{2}}(1 + i\alpha)(N g_{1} - \gamma)E_{1} +  \eta_1 \label{eq:1}  \\
&&\dot{E}_{2} = \textstyle{\frac{1}{2}}(1 + i\alpha)(N g_{2} - \gamma)E_{2} +  \eta_2 \label{eq:2} \\
&&\dot{N} = J_{0} - \frac{N}{\tau_{s}} - N \sum_{m}g_{m}|E_{m}|^2, \label{eq:3} 
\end{eqnarray}
where the nonlinear modal gain is given by
\begin{eqnarray}
& g_{m} = g_{m}^{(0)}\left(1 + \epsilon\sum_{n}\beta_{mn}|E_{n}|^2\right)^{-1}. \label{eq:4}
\end{eqnarray}
In the above equations, $g_{m}^{(0)} = 1$ is the linear modal gain. $J_{0}$ is the pump current, 
$\gamma$ is the cavity decay rate, and $\tau_s$ is the carrier lifetime. The phase-amplitude
coupling is given by $\alpha$. The parameters $\epsilon \beta_{mn}$ determine 
the cross and self saturation of the gain. A stable two-mode solution in the free running 
laser requires $\beta_{mn} < \beta_{mm}$. The Gaussian white noise terms $\eta_m(t)$ fulfill
the relation $\langle \eta_m(t) \eta_n(t')\rangle = D \delta(t-t') \delta_{mn}$. A similar 
model has been successfully used to describe the dynamical features of two-mode lasers with 
optical injection \cite{osborne_09}.

Using the standard theory of multivariate Ornstein-Uhlenbeck processes
\cite{gardiner_09} we obtain the spectrum matrix in the small noise
limit at the two colour point by using the general formula
\begin{equation}
S(\omega) = \frac{1}{2\pi} (J + I i \omega)^{-1} \hat{D} (J^T - I i \omega)^{-1}. \label{eq:6}
\end{equation}
Here $J$ is the Jacobian matrix of the system
\ref{eq:1} - \ref{eq:3} evaluated at the two-color point and
$\hat{D} = \text{diag}(D,D,0)$ is the noise coupling matrix. The
resulting noise spectra for the individual modes and for the total
output are shown as dashed lines in Fig.~\ref{2mspc4A} (b). One can see that  
we obtain good qualitative agreement with the corresponding experimental data, 
including the characteristic feature of the antiphase noise peak at low 
frequencies of the individual modes and its absence in the total output. 
The analytical treatment reveals that the origin of the antiphase noise in
the dynamics of the individual modes can be traced to one negative real
eigenvalue of the Jacobian, whose eigenvector corresponds to
fluctuations around the two-color point with a fixed total output and
a fixed carrier density. On the other hand, the relaxation oscillation 
peak at around $5\;$GHz, which is visible in the spectra for the total and
the individual modes, is related to a pair of complex conjugated eigenvalues 
of the Jacobian that describe the exchange of energy between the carriers
and the fields. For these calculations the value of $J_{0}$ was fixed at 
75\% above its threshold value, $J_{\mbox{\footnotesize thr}}$. Other parameters 
were $(\gamma \tau_s) = 800$, $\epsilon = 0.01$, and 
$\beta \equiv \beta_{mn}/ \beta_{mm} = 2/3$.

\begin{figure}
\centering
\includegraphics[width=8.0cm]{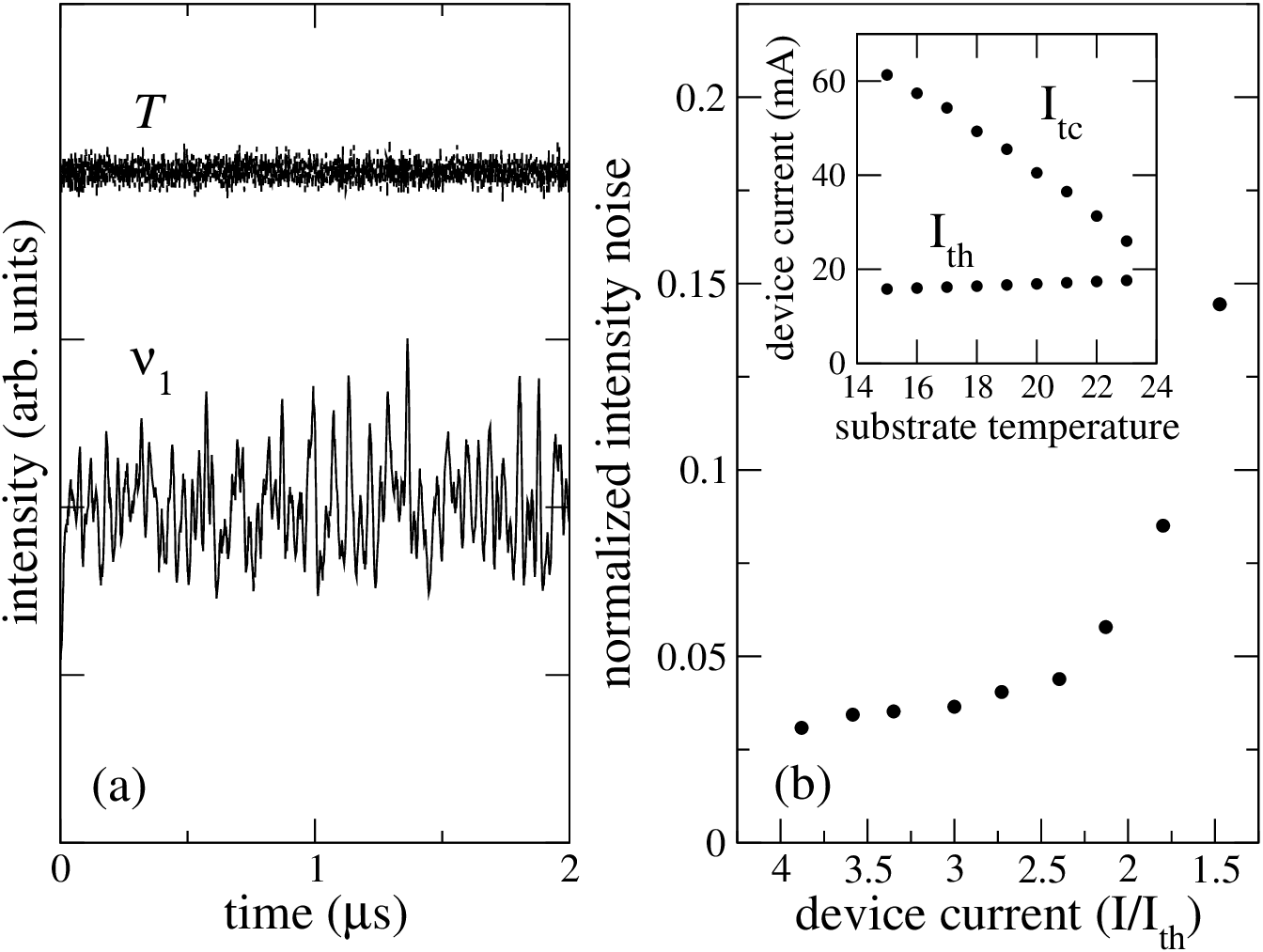}
\caption{\label{2mspc4B} (a) Intensity time traces of one of the primary modes, $\nu_1$, and 
of the total intensity, $T$, measured at the two-color point. (b) Variation of the normalized 
standard deviation of the intensity noise in one of the primary modes measured at the 
two-color point. Inset: Variation of the device current measured at threshold, 
$\mbox{I}_{\mbox{\scriptsize th}}$, and at the two-color point, 
$\mbox{I}_{\mbox{\scriptsize tc}}$, as the substrate temperature is varied.}
\end{figure}

Time traces of the total intensity and of one of the primary modes that illustrate the 
enhanced antiphase noise are shown in Fig. \ref{2mspc4B} (a). On this scale the total
intensity is almost constant, while large fluctuations are visible in the
intensity of the primary mode. These intensity fluctuations in the individual 
modal intensities will lead to increased intensity noise in a generated THz signal. 
However, we would like to point out that because the instantaneous frequency of each 
mode is determined by the dynamics of the carrier density, which in turn depends on 
the dynamics of the total intensity, antiphase noise does not lead to an additional 
increase in the phase noise of the THz signal. 

We have found that the level of the intensity noise is strongly dependent on how far the 
two-color point is located above the device threshold. By varying the substrate temperature, 
the position of the gain peak can be moved and the two-color point can be brought further from 
threshold. The normalised standard deviation of the
intensity noise is plotted in Fig. \ref{2mspc4B} (b). One can see that the increased 
optical power leads to a reduction in the level of intensity noise to less than 5\%.
A device similar to that shown in Fig. \ref{2mspc4A} driven at five times threshold was 
used to generate THz radiation in a photomixing setup \cite{osborne_08}. 

Because of the increased efficiency of photomixing and applications in high bandwidth wireless
communications, the generation of beat notes in the mmWave region around 100 GHz is of
significant interest. However, we have found that the simple two-mode device cannot be used 
in this region because of the phenomenon of wavelength bistability. Wavelength bistability
is expected to appear in quantum-well semiconductor lasers once the cross-saturation
of the gain exceeds the self-saturation \cite{obrien_06a}. We have found that this boundary 
is located in the region of 300 GHz. For primary mode spacings less than this, a discontinuous
switching of the intensity from one primary mode to the other is observed over a very small 
current range. 

\begin{figure}
\centering
\includegraphics[width=8.0cm]{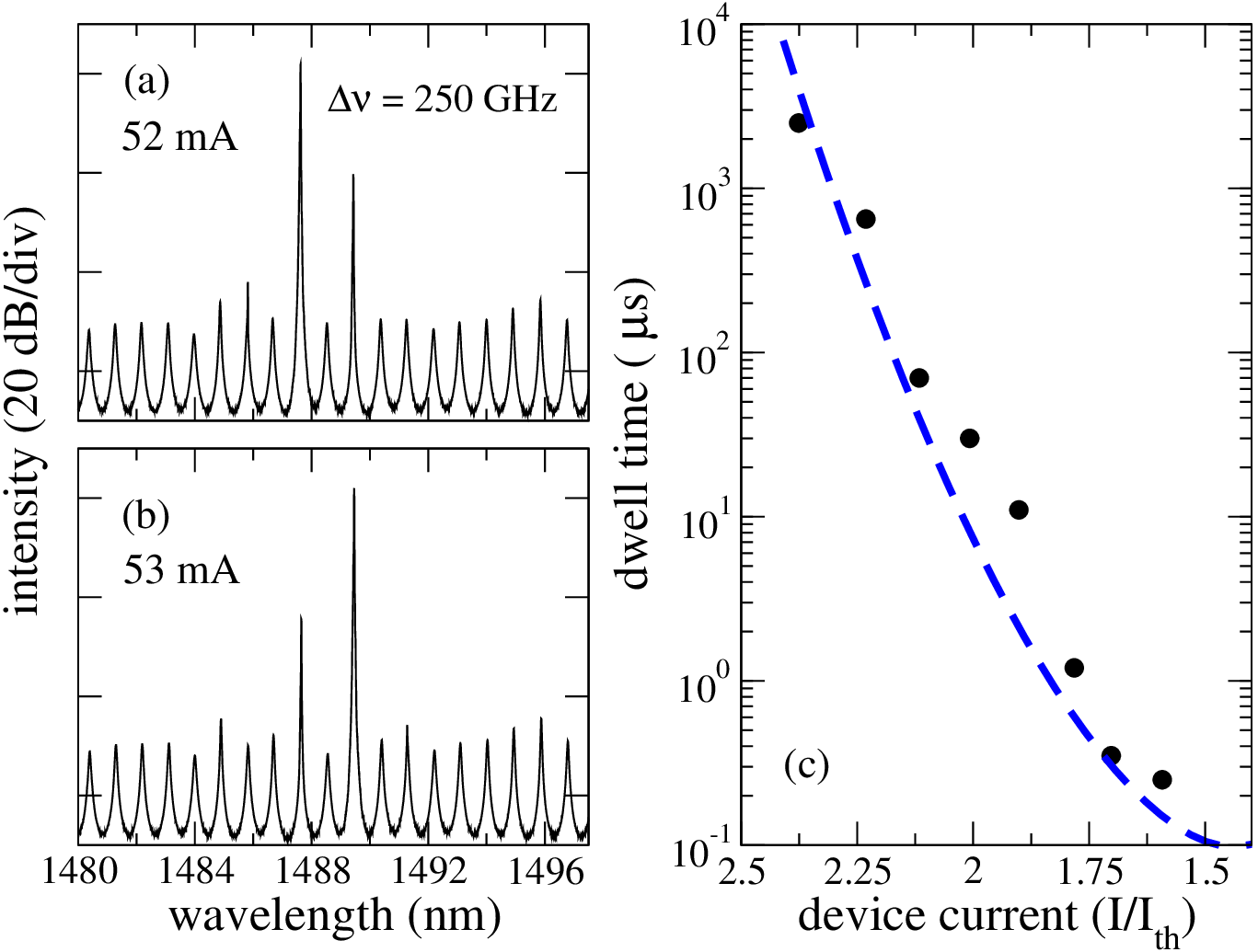}
\caption{\label{2mspc2} Optical spectra of the wavelength bistable device before (a) 
and after (b) the bistability region. (c) Variation of the dwell time in the bistability 
region as the device current at the bistability point is varied. The
dashed line is obtained from the escape time analysis of a reduced
one-dimensional model [Eqn. \ref{eq:8}].}
\end{figure}

Optical spectra that illustrate this switching are shown in
Fig. \ref{2mspc2}.  These are experimental measurements from a
two-mode device with a peak emission near 1490 nm. The primary mode
spacing in this case is 250 GHz and one can see that a 20 dB change in
intensity is observed over approximately 1 mA variation in the device
current. In fact, these quasi-single mode states coexist over a small
hysteresis region, within which noise driven switching of the
intensity between the modes can be observed. This so-called
mode-hopping behaviour \cite{ohtsu_89} is associated with a
characteristic dwell time, which is the average time measured between
switching events. By varying the substrate temperature as before we
were able to move the bistability region further from threshold. We
found that the corresponding dwell time varied by over four orders of
magnitude as shown in Fig. \ref{2mspc2} (c). The dashed line in
Fig.~\ref{2mspc2} (c) shows a theoretical calculation of the dwell
time, which is obtained by approximating the system
\ref{eq:1} - \ref{eq:3} by a one dimensional stochastic
differential equation for the dynamical variable
$\phi=\arctan\left(\left|E_1\right|^2 /\left|E_2\right|^2 \right)$,
similar to the approach which was taken in \cite{pedaci_06}. This
leads to an Arrhenius formula for the dwell time $\tau_{\mbox{\tiny D}}$ given by
\begin{equation}
  \label{eq:8}
  \tau_{\mbox{\tiny D}} = \frac{\sqrt{2}\pi}{\sqrt{e D \gamma \tau_s \delta}
    \left(\hat{P}^{2}-1\right)}\exp \left[\textstyle{\frac{1}{2}}\hat{P}^{2}\right] ,
\end{equation}
where $\hat{P} = P\sqrt{\gamma \delta /D\tau_s}$ with 
$P = \textstyle{\frac{1}{2}}\left(\frac{J_0}{J_{\mbox{\scriptsize thr}}} - 1\right)$ 
and $\delta = \textstyle{\frac{1}{2}}\epsilon \left(\beta - 1 \right)$. For this calculation,
we used the same set of parameters as in Fig.~\ref{2mspc4A} (b) apart from the ratio of
cross to self saturation which is now given by $\beta = 1.1$. The noise 
strength used for fitting experimental data was $D=9.6 \times 10^{15} \mbox{ s}^{-2}$.

\section{Millimeter-wave frequency synthesis with passively mode-locked discrete mode lasers}

In order to avoid the problems associated with wavelength bistability in two-mode 
quantum-well devices, we have designed passively mode-locked devices that support a 
minimal set of four primary modes. These devices include a saturable absorber section 
placed adjacent to one of the cavity mirrors. In the device we consider here, 
four primary lasing modes are chosen with a spacing of two fundamental modes leading 
to mode-locking at the first harmonic of the cavity. 

\begin{figure}
\centering
\includegraphics[width=8.0cm]{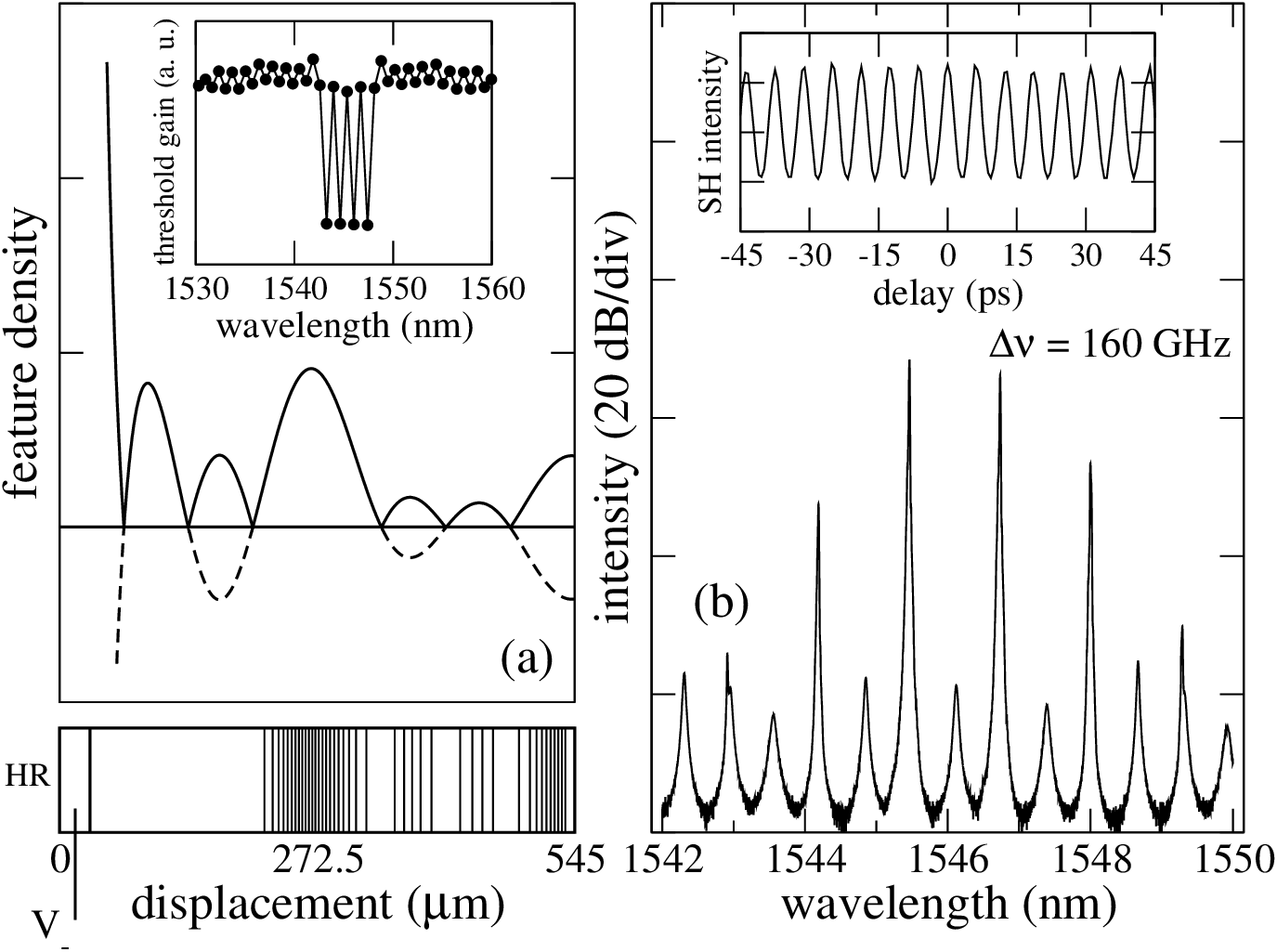}
\caption{\label{4mde_desgn} (a) Feature density function (solid line). Dashed lines 
indicate intervals where the Fourier transform of the spectral filtering function 
chosen is negative. Inset:  Calculation of the threshold gain of 
modes for the laser cavity schematically pictured in the lower panel of the figure. 
Lower panel: Laser cavity schematic indicating the locations of the additional features. 
The device is high-reflection (HR) coated and includes a saturable absorber section as 
indicated. (b) Optical spectrum of the device at a device current of 110 mA and
a reverse bias voltage of -2.8 V. Inset: Background free intensity autocorrelation
measurement showing mode beating at 160 GHz.}
\end{figure}

The cavity design in this case represents a straightforward extension to four modes of 
the approach to designing a two-mode device described in \cite{obrien_06a}. Fig. 
\ref{4mde_desgn} (a) shows the feature density function which must
be sampled in order to define the effective index profile. A schematic picture of the 
device, high-reflection coated as indicated, is shown in the 
lower panel of Fig. \ref{4mde_desgn} (a). The device length is 545 $\mu$m, and $N = 42$ etched
features are introduced. The calculated form of the threshold gain spectrum is shown in the 
inset of Fig. \ref{4mde_desgn} (a). 

Experimental measurements of a ridge waveguide Fabry-Perot laser fabricated to the 
design depicted are shown in Fig. \ref{4mde_desgn} (b). This device has a peak emission near 
1.5 $\mu$m and a saturable absorber section of length 30 $\mu$m was placed adjacent to the high 
reflectivity mirror. Fig. \ref{4mde_desgn} (b) displays the optical spectrum for a reverse bias 
voltage of -2.8 V applied to the saturable absorber section. The four selected modes are in excess
of 20 dB stronger than all other modes and have a frequency separation of 160 GHz. 
Because most of the laser light is concentrated in the two central primary modes, the 
resulting time evolution of the power is essentially a beating between these two modes. 
The inset of Fig. \ref{4mde_desgn} (b) displays the intensity autocorrelation which confirms
a sinusoidal intensity modulation at the mode-locking frequency. 

We note that while examples of mode-locked diode lasers with a sinusoidal intensity modulation
have been presented in the literature \cite{avrutin_02}, the sinusoidal modulation can be attributed 
to the fact that the primary mode spacing is comparable to the gain bandwidth in these devices. For 
narrower mode spacings a mode locked device will generally oscillate on many modes generating pulsed output.
External filtering of pairs of modes from the mode-locked spectrum is then necessary to obtain sinusoidal
output \cite{pelusi_97}. Our approach to this problem can be regarded as integrating the functions of
mode-locking and spectral filtering in a single device. 

These results demonstrate that discrete mode devices provide a convenient means to
realize THz and mmWave frequencies in the optical domain. Microwave frequency synthesis 
will also be possible in longer devices with a correspondingly narrower mode spacing. 
We are currently investigating the dependence of the linewidth of the beatnote on 
the drive parameters. Preliminary measurements indicate that in the mode-locked device, 
this linewidth is in the region of 5 MHz. It should be noted however that because the mode 
selection mechanism in our device acts to predetermine the carrier wave frequency in the 
mode-locked state, we expect that the characteristics of this state will also depend on 
where the central mode frequency is placed with respect to the material gain and loss 
dispersion. 

While mode-locking was achieved at 160 GHz in the current device, results
suggest that quantum-well material can be used for mode-locking frequencies as large 
as 1 THz \cite{arahira_94}. However, the use of self-assembled gain materials may provide 
significant advantages such as reduced amplified spontaneous emission noise, faster 
carrier recovery and limited alpha factors that can further improve the stability 
of the mode-locked state. The possibilities suggested by techniques such as hybrid 
mode-locking are also interesting in this regard. Significantly, single-mode discrete 
mode devices have been shown to have optical linewidths as small as 100 kHz under 
normal drive conditions \cite{kelly_07}. This suggests that a mode beating linewidth 
in the sub-MHz range should be achievable with discrete mode devices. 

\section{Conclusion}

We have demonstrated that discrete mode diode lasers provide a direct route to
optical synthesis of THz and mmWave frequencies. These devices are specially designed
multimode semiconductor lasers with a pre-defined number and spacing of primary 
lasing modes. For frequency separations in the THz region, it was shown that a 
single section quantum-well device can support two optical modes simultaneously with 
high spectral purity. We showed that wavelength bistability appears for mode spacings 
less than approximately 300 GHz and that this phenomenon prevents simultaneous 
lasing of two modes. Finally, we showed that passive mode-locking of devices that 
support only four primary modes can be used to overcome the difficulties 
presented by wavelength bistability. In this case sinusoidal modulation of the optical 
intensity at mmWave frequencies was achieved.  

\section*{Acknowledgment}

The authors thank Science Foundation Ireland and Enterprise Ireland for financial
support of this work.


\begin{thebibliography}{99}

\bibitem{yao_09}
J.~Yao, ``Microwave photonics,'' J. Lightwave Tech., vol. 27, no. 3, pp. 314-335, Feb. 2009.
\bibitem{siegel_02}
P.~H.~Siegel, ``Terahertz technology,'' IEEE Trans. Microwave Theory Tech., vol. 50, no. 3, 
pp. 910-928, Mar. 2002.
\bibitem{steele_83}
R.~C.~Steele, ``Optical phase-locked loop using semiconductor laser diodes,'' 
IEE Electron. Lett., vol. 19, no. 2, pp. 69-71, Jan. 1983.
\bibitem{friederich_10}
F.~Freiderich, G.~Schuricht, A.~Deninger, F. Lison, G.~Spickermann, P.~H.~Bol\'ivar, and
H.~G.~Roskos, ``Phase locking of the beat signal of two distributed-feedback diode lasers to oscillators
working in the MHz to THz range,'' Opt. Express, vol. 18, no. 8, pp. 8621-8629, Apr. 2010.
\bibitem{mcintosh_95}
K.~A.~McIntosh, E.~R.~Brown, K.~B.~Nichols, O.~B.~McMahon, W.~F.~DiNatale, 
and T.~M.~Lyszczarz, ``Terahertz photomixing with diode lasers in low-temperature-grown GaAs,'' 
Appl. Phys. Lett., vol. 67, no. 26, pp. 3844-3846, Dec. 1995.
\bibitem{hirata_03}
A.~Hirata, M.~Harada, and T.~Nagatsuma, ``120-GHz wireless link using photonic techniques for generation,
modulation, and emission of millimter wave signals,'' J. Lightwave Tech., vol. 21, no. 10, pp. 2145-2153, Oct. 2003.
\bibitem{fukushima_03}
S.~Fukushima, C.~F.~C.~Silva, Y.~Muramoto, and A.~J.~Seeds, ``Optoelectronic millimeter-wave synthesis 
using an optical frequency comb generator, optically injection locked lasers, and a unitraveling-carrier
photodiode,'' J. Lightwave Tech., vol. 21, no. 12, pp. 3043-3051, Dec. 2003.
\bibitem{tani_05}
M.~Tani, O.~Morikawa, S.~Matsuura, and M.~Hangyo, ``Generation of terahertz radiation by
photomixing with dual- and multiple-mode lasers,'' Semicond. Sci. Technol., vol. 20, no. 7, pp. 151-163, Jul. 2005.
\bibitem{park_04}
I.~Park, I.~Fischer and W.~Els\"a\ss er, ``Highly nondegenerate four-wave mixing in a tunable dual-mode 
semiconductor laser,'' Appl. Phys. Lett., vol. 84, no. 25, pp. 5189-5191, Jun. 2004.
\bibitem{iio_95}
S.~Iio, M.~Suehiro, T.~Hirata, and T.~Hidaka, ``Two-longitudinal-mode laser diodes,'' IEEE Photon. Tech. 
Lett., vol. 7, no. 9, pp. 959-961, Sep. 1995.
\bibitem{matsui_99} 
Y.~Matsui, M.~D.~Pelusi,  S.~Arahira, and Y.~Ogawa, ``Beat frequency generation 
up to 3.4 THz from simultaneous two-mode lasing operation of sampled-grating DBR laser,'' IEE Electron. 
Lett., vol. 35, no. 6, pp. 472-474, Mar. 2008.
\bibitem{roh_00}
S.~D.~Roh, T.~S.~Yeoh, R.~B.~Swint, A.~E.~Huber, J.~S.~Woo, J.~J.~Coleman, ``Dual-wavelength 
InGaAs-GaAs ridge waveguide distributed Bragg reflector lasers with tunable mode separation,''
IEEE Photon. Technol. Lett., vol. 12, no. 10, pp. 1307-1309, Oct. 2000. 
\bibitem{avrutin_02}
D.~A.~Yanson, M.~W.~Street, S.D.~McDougall, I.~G. Thayne, J.~H.~Marsh, and E.~A.~Avrutin, ``Ultrafast 
harmonic mode-locking of monolithic compound-cavity laser diodes incorporating photonic-bandgap 
reflectors,'' IEEE J. Quantum Electron., vol. 38, no. 1, pp. 1-11, Jan. 2002.
\bibitem{kim_09}
N.~Kim, J.~Shin, E.~Sim, C.~W.~Lee, D.-S.~Yee, M. Y.~Jeon, Y.~Jang, and K.~ H.~Park, ``Monolithic dual-mode 
distributed feedback semiconductor laser for tunable continuous-wave terahertz generation,'' Opt. Express, 
vol. 17, no. 16, pp. 13851-13859, Jul. 2009. 
\bibitem{obrien_05}
S.~O'Brien and E.~P.~O'Reilly, ``Theory of improved spectral purity in index patterned Fabry-Perot lasers,'' 
Appl. Phys. Lett., vol. 86, no. 20, 201101, May 2005.
\bibitem{obrien_06a}
S.~O'Brien, S.~Osborne, K.~Buckley, R.~Fehse, A.~Amann, E.~P.~O'Reilly, L.~P.~Barry, 
P.~Anandarajah, J.~Patchell, and J.~O'Gorman, ``Inverse scattering approach to multiwavelength 
Fabry-Perot laser design,'' Phys. Rev. A, vol. 74, no. 6, 063814, Dec. 2006.
\bibitem{kelly_07}
B.~Kelly, R.~Phelan, D.~Jones, C.~Herbert, J.~O'Carroll, M.~Rensing, J.~Wendelboe, C.~B.~Watts, 
A.~Kaszubowska-Anandarajah, P.~Perry, C.~Guignard, L.~P.~Barry, and J.~O'Gorman, ``Discrete mode laser
diodes with very narrow linewidth emission,'' IEE Electron. Lett., vol. 43, no. 23, pp. 1282-1284, Nov. 2007.
\bibitem{osborne_09}
S.~Osborne, A.~Amann, K.~Buckley, G.~Ryan, S.~P.~Hegarty, G.~Huyet, and S.~O'Brien, 
``Antiphase dynamics in a multimode semiconductor laser with optical injection,'' 
Phys. Rev. A, vol. 79, no. 2, 023834, Feb. 2009.
\bibitem{gardiner_09}
C.~Gardiner, \emph{Stochastic Methods}, Springer, Berlin, 2009.
\bibitem{pedaci_06}
F.~Pedaci, S.~Lepri, S.~Balle, G.~Giacomelli, M.~Giudici, and J.~R. Tredicce, ``Multiplicative 
noise in the longitudinal mode dynamics of a bulk semiconductor laser,'' Phys.~Rev.~E, vol.~73, 
no. 4, 041101, Apr. 2006.
\bibitem{osborne_08}
S.~Osborne, S.~O'Brien, E.~P.~O'Reilly, P.~G.~Huggard, and B.~N.~Ellison, ``Generation of CW 0.5 THz
radiation by photomixing the output of a two-colour 1.49 $\mu$m Fabry-Perot laser diode,'' IEE Electron. 
Lett., vol. 44, no. 4, pp. 296-297, Feb. 2008.
\bibitem{ohtsu_89}
M.~Ohtsu and Y.~Teramachi, ``Analyses of mode partition and mode hopping in 
semiconductor lasers,'' IEEE J. Quantum Electron., vol. 25, no. 1, pp. 31-38, Jan. 1989.
\bibitem{pelusi_97}
M.~D.~Pelusi, H.~F.~Liu, D.~Novak, and Y.~Ogawa, ``THz optical beat frequency generation from a single
mode locked semiconductor laser,'' Appl. Phys. Lett., vol. 71, no. 4, pp. 449-451, Jul. 1997.
\bibitem{arahira_94}
S.~Arahira, S.~Oshiba, Y.~Matsui, T.~Kunii, and Y.~Ogawa, ``Terahertz-rate optical
pulse generation from a passively mode-locked semiconductor laser
diode,'' Optics Lett., vol. 19, no. 11, pp. 834-836, Jun. 1994.

\end{thebibliography}
\end{document}